\documentclass[11pt]{article}
\usepackage{amssymb,amsmath}
\usepackage{graphicx}
\usepackage{epsfig}
\usepackage{color}

\input{tcilatex}

\begin{document}

\title{Does Reactor Neutrino Experiment Play an Important
  Role in $\theta_{13}$ of Lepton Mixing (PMNS) Matrix ? }  

\date{}
\author{
{\large Q. Y. Liu$^{1,2}\footnote{email: qiuyu@ustc.edu.cn}$, J. Deng$^{1}$, B.L. Chen$^{1}$, P. Yang$^{1}$}
\bigskip
\\
{
$^{1}$ {\small \it Department of Modern Physics, University of Science
    and}
}
\\
{
{\small \it  Technology  of China, Hefei, Anhui 230026, China.}
}
\\
{
$^2$ {\small \it The Abdus Salam International Center for
Theoretical Physics,}
}
\\
{
{\small \it Strada Costiera 11,  34100, Trieste,
Italy.}
}
}
\maketitle

\begin{abstract}
Reactor neutrinos play an important role in determining 
parameter $\theta_{13}$ in the lepton mixing (PMNS) matrix. A next
important step on measuring PMNS matrix
could be to build another reactor neutrino experiment, for example, 
DaYa Bay in China, to search the possible oscillations via $\sin^2 2
\theta_{13}$ and $\Delta m^2_{13}$. We consider 4 different schemes for
positions of three 8-ton detectors 
of this experiment, and simulate the results with respect to an array 
of assumed ''true'' values of physics parameters. Using three kinds of analysis
methods, we suggest a best scheme for this experiment  which is to place
a detector $2200m \sim 2500m $ symmetrically away from two reactors, and
to put the other two detectors closer to their corresponding
reactors respectively, almost at a $100m \sim 200m$ distance.
Moreover, with conservative assumption on the experimental
technique, we construct series of allowed 
regions from our simulation results, and give detailed explanations 
therein.      
\end{abstract}

\vspace{3mm}


\vskip 5cm {\large \textbf{ 
}} \vfill \eject
\baselineskip=0.36in \renewcommand{\theequation}{\arabic{section}.%
\arabic{equation}} \renewcommand{\thesection}{\Roman{section}} 
\makeatletter
\@addtoreset{equation}{section} \makeatother       

\section{ Introduction }
\bigskip
An next step in
the exciting field of neutrino physics would be to improve current measurements and to
measure some of the remaining unknown parameters in the full $3 \times
3$ leptonic
flavor mixing, which is called the PMNS mixing matrix. There are
important
differences between the PMNS and quark CKM matrices. Which may be 
essential for our 
understanding the underlying physics. In addition to three masses $m_{i}$, there
are 6 free parameters in the matrix. We may parameterize $U_{PMNS}$
\cite{PMNS} as follows:  
\begin{equation}
U_{PMNS}=U_{12}\times U_{23}\times U_{13}\times U_{major}.
\end{equation}%
\begin{eqnarray}
U_{12} &=&\left( 
\begin{array}{ccc}
\cos \theta _{12} & \sin \theta _{12} & 0 \\ 
-\sin \theta _{12} & \cos \theta _{12} & 0 \\ 
0 & 0 & 1%
\end{array}%
\right) \\
U_{23} &=&\left( 
\begin{array}{ccc}
1 & 0 & 0 \\ 
0 & \cos \theta _{23} & \sin \theta _{23} \\ 
0 & -\sin \theta _{23} & \cos \theta _{23}%
\end{array}%
\right) \\
U_{13} &=&\left( 
\begin{array}{ccc}
\cos \theta _{13} & 0 & e^{-i\delta _{CP}}\sin \theta _{13} \\ 
0 & 1 & 0 \\ 
-e^{-i\delta _{CP}}\sin \theta _{13} & 0 & \cos \theta _{13}%
\end{array}%
\right) \\
U_{major} &=&\left( 
\begin{array}{ccc}
1 & 0 & 0 \\ 
0 & e^{-i\alpha } & 0 \\ 
0 & 0 & e^{-i\beta }%
\end{array}%
\right)
\end{eqnarray}

\bigskip

\bigskip

$U_{23}$ and the value of $\Delta m_{23}^{2}$ have been measured by the
Super-Kamiokande \cite{SK_atm} and K2K long base line experiments
\cite{K2K}; while $U_{12}$ and the value 
of $\Delta m_{12}^{2}$ are parameters of the confirmed solar neutrino MSW 
\cite{MSW} solution. $U_{major}$ is the possible majorana phase matrix, its
values and the overall mass scale will require kinematical and neutrino-less
double-$\beta $ decay measurements. $U_{13}$ is the next goal for the
experiments since it will tell us, in particular, if $CP$ violation is possible in the lepton
sector. We can expect it from the very long base line experiments such
as  JPARC, H2B
\cite{H2B,LBL} and, new generation of reactor neutrino experiments.

\bigskip 
The CHOOZ experiment \cite{CHOOZ,threeNu} only gives an upper limit on the
mixing angle $\theta_{13}$ ($\sin^{2}2\theta_{13}<0.10$ at $90\%$ 
confidence level). There are attempts to find this mixing
angle in LBL accelerator experiments \cite{H2B,LBL}, or in three
neutrino analysis of solar neutrino \cite{LBL,3nusun,global}, but the precision is very difficult to
achieve. Since $\Delta m_{12}^{2}\ll \Delta m_{23}^{2}$, it must happen that 
$\Delta m_{13}^{2}\approx \Delta m_{23}^{2}$.  A new generation of reactor
experiments has been proposed to search for $\bar{\nu}_{e}$ disappearance at
baselines of ~1 $\sim$ 2 km corresponding to this value of $\Delta
m^{2}$. To improve 
on the mixing angle sensitivity achieved by Palo Verde and CHOOZ, proposals
for reactor $\theta _{13}$ experiments include a large detector to reduce
the statistical error, and also a second detector positioned very close ($%
\sim $  100 m) to the reactor. The near detector would precisely measure the
incident flux, providing to drop out many systematic uncertainties in
the flux calculation.  This also requires the detectors to be made
identical and/or movable. Sensitivity down to $\sin ^{2}2\theta _{13}\approx 10^{-2}$ seems
within grasp. Such experiments were discussed in literature 
\cite{Mikaelyan, Wang1}. A practical possibility is a reactor experiment at
DaYa-Bay, which is located near a special economic zone in Guang-Dong
Province in southern China. There are nuclear power plants in that area.

The knowledge we have about neutrino mixing is powerful
to judge Grand Unified models, such as the most inspiring
SO(10) GUT. Starting from the lepton quark symmetry in this model, one is
able to obtain a bi-large neutrino mixing pattern via see-saw
mechanism. It leads to a non-zero $\sin^2 2 \theta_{13}$, e.g., at about
$0.09$ in paper \cite{so10}, which is out of CHOOZ's limit but is very
easy to
discover in DaYa-Bay experiment within one year operation.   

\begin{figure}[htbp]
\vspace*{-0.6cm} 
\centerline{\epsfxsize=8cm \epsfysize=8cm
\epsfbox{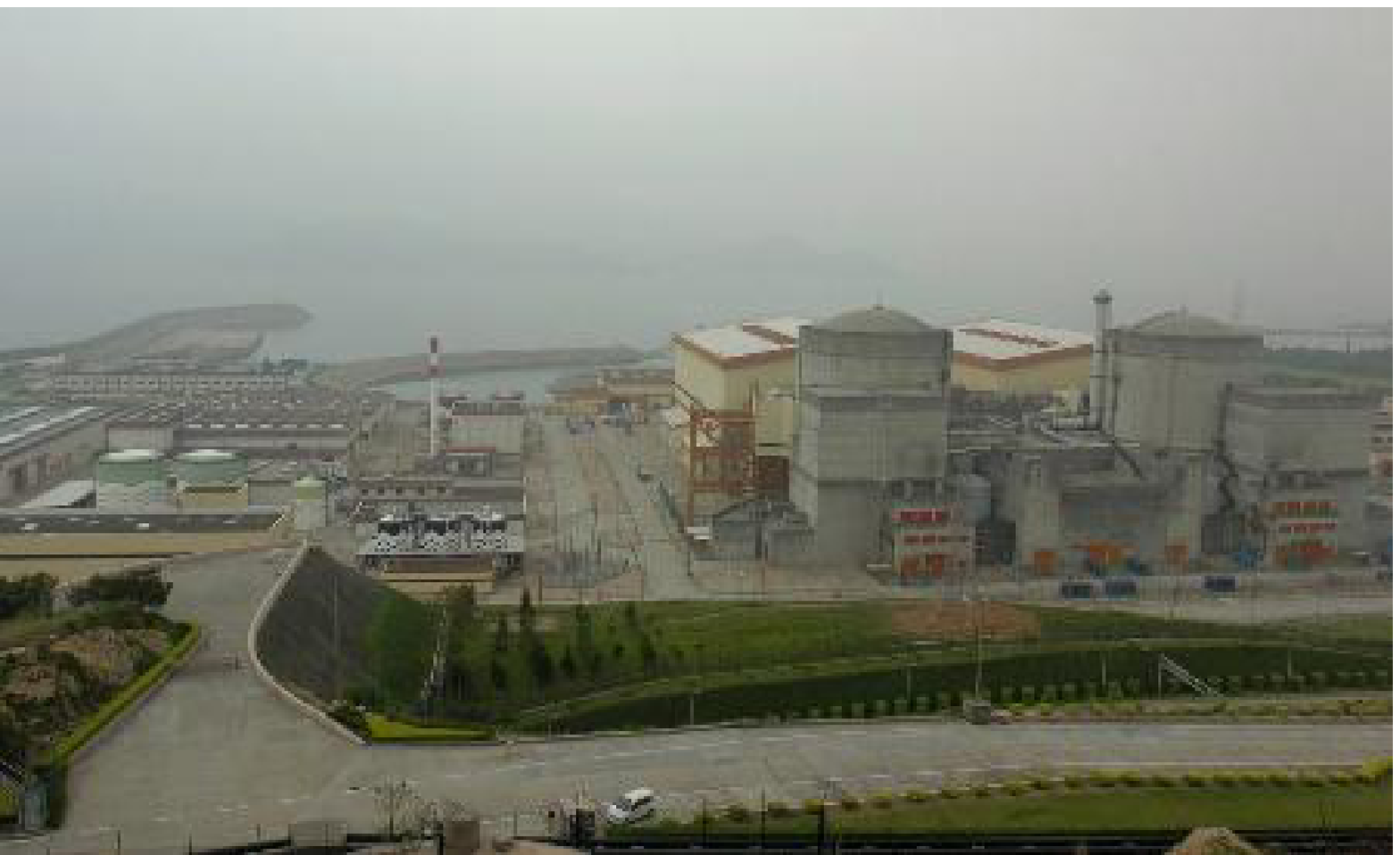}}
\caption{The picture of the DaYa-Bay reactor. }
\label{DaYaoutlook}
\end{figure}

In this paper we will describe the importance of reactor neutrinos
in determination of $\sin^2 2\theta_{13}$ - $\Delta m^2_{13}$; we will
concentrate on 
 possible new China experiment and its goal - finding $\sin^2
2\theta_{13}$. The paper is organized as follows: first, we will
describe a
possible reactor experiment in DaYa-Bay (see 
fig. \ref{DaYaoutlook}), which is a kind of upgraded 
CHOOZ experiment\cite{Mikaelyan,Minakata,Huber}; its possible
systematic and statistical uncertainties are analogies to the
CHOOZ's one. Next we will explain different methods of analysis, and
importantly,  our arrangement of the detectors' positions, accompanied
by our simplified Monte Carlo simulations on the possible sensitivity
regions for $\sin^2 2 \theta_{13}$; and discovery potentialities are
discussed therein. Finally, we will give our conclusion of the paper. 

\section{Reactor Neutrinos}

\bigskip

In a reactor, anti-neutrinos are released by radioactive isotope
fission; the total neutrino spectrum is a rather well understood function of
the thermal power $W$, the amount of thermal power $w_{i}$ emitted during
the fission of a given nucleus, and the isotopic composition of the reactor
fuel $f_{i}$, 
\begin{equation}
S(E_{\nu })=\frac{W}{\sum f_{i}w_{i}}\sum f_{i}\left( \frac{dN}{dE_{\nu }}%
\right) _{i}
\end{equation}
The index $i$ of $f_{i}$ stands for 4 isotopes such as $^{235}U$,
 $^{238}U$,  $^{239}Pu$ and $^{241}Pu.$ The (dN/dE) is the
energy spectrum of the fissionable isotope, it can be parameterized by the
following expression\cite{nu_spectrum} when $E_{\nu }\geq $ $2MeV$: 
\begin{equation}
\frac{dN_{\nu }}{dE_{\nu }}=e^{a_{0}+a_{1}E_{\nu }+a_{2}E_{\nu }^{2}}
\end{equation}%
the coefficients $a_{i}$ depend on the nature of the fissionable isotope.
KamLAND is a scintillator detector, where electronic anti-neutrinos are
detected by free protons via inverse $\beta -$decay reaction\cite%
{nu_spectrum}, 
\begin{equation}
\overline{\nu }_{e}+p\rightarrow e^{+}+n  
\label{inverse-decay}
\end{equation}%
In the limit of infinite nucleon mass, the cross section of this reaction is
given by $\sigma (E_{\nu })=kE_{e^{+}}P_{e^{+}}$, where $%
E_{e^{+}},~P_{e^{+}} $ are the positron energy and momentum respectively and 
$k$ can be taken as $k=9.55\times 10^{-44}~cm^{2}/MeV^{2}$. The anti-neutrino
events are characterized by the positron annihilation signal and the delayed
neutron capture sign \cite{KamLAND}. 

From the reactor to the detector, massive neutrinos oscillate on the way and
change their flavor composition to a certain extent. The anti-neutrino $%
\overline{\nu }_{e}$ can oscillate to other flavors via $\Delta m_{12}^{2}$
and $\Delta m_{13}^{2}$. 
In principle, reactor neutrinos are able to give us information about $\Delta
m_{13}^{2}$ $\&$ $\Delta m_{12}^{2}$ and two mixing angles $\sin ^{2} 2\theta
_{13}$ $\&$ $\sin ^{2}2\theta _{12}$, which are almost all the neutrino
oscillation parameters except $\sin ^{2} 2\theta _{23}$. This
advantage of reactor neutrinos is due to its low energy, which is several MeV,
comparing with accelerator neutrinos' GeVs. Reactor experiment like DaYa Bay
 is sensitive to $\sin ^{2}2\theta _{13}$
and $\Delta m_{13}^{2}$. This is the second order oscillating
effect of the reactor neutrinos, comparing with the oscillations induced by 
$\theta _{12}$ and $\Delta m_{12}^{2}$. Since the energy of detectable
reactor neutrino is about MeVs, from atmospheric oscillation
experiments, $\Delta m_{13}^{2}$ is in the range of $1.3\sim 5.0\times
10^{-3}\mathtt{eV}^{2}$ at $99\%$C.L. \cite{SK_atm}, with the best fitted
point $2.5\times 10^{-3}\mathtt{eV}^{2}$, so the expected maximum
oscillation is reached at about 1500 m. Then the survival probability
is reduced to the two flavor neutrino case:  
\begin{equation}
P(E_{\nu },L,\theta _{13},\Delta m_{13}^{2})=1-\sin ^{2}(2\theta _{13})\sin
^{2}\left( \frac{1.27\Delta m_{13}^{2}(\mathtt{eV}^{2})L(\mathtt{m})}{E_{\nu
}(\mathtt{MeV})}\right) 
\end{equation}
where $\sin ^{2}2\theta _{13}$ is the only unknown parameter here,
which is our major interest in this paper.

\bigskip

\section{Possible \protect\bigskip  Reactor Neutrino Experiment in
  China }

\bigskip DaYa-Bay is one of the Chinese running reactor power plants, located in Guang-Dong province, south China, and it is quite close to Hong
Kong. The DaYa Bay nuclear power plant consists of two twin reactor cores,
separated about $1200$ m from each other, one is called DaYa, the other is
called LingAo, each core can generate a thermal power of $2.9$ $GW$, making a
total of $11.6$ $GW$, and a third twin core is planned to be on line in 2010.
The reactors are built near a mountain, so it is possible to build a near
experiment hall with an overburden of $400$ $mwe$ at a distance of
about $300$ $m$
to the core and a far hall with an overburden of $1200$ $mwe$ at a
distance of about $1500\sim 2000$ $m$ to the core. This is a certain
improvement in comparison with the
 CHOOZ experiment which has a value of $8.5$ $GW$ in thermal power, and 
$300$ $mwe$ in rock overburden.

\begin{figure}[htbp]
\vspace*{-0.6cm} 
\centerline{\epsfxsize=8cm \epsfysize=8cm
\epsfbox{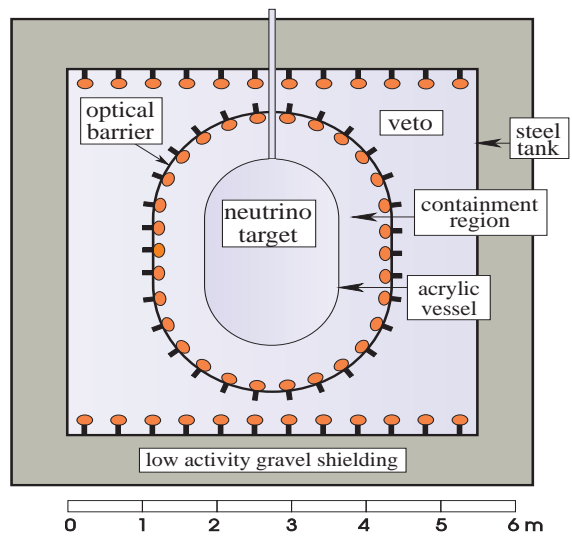}}\vspace*{-0.6cm}
\caption{Structures of a detector for reactor neutrino experiment. the
 experiment like DaYa-Bay   
 can have three such 8-ton detectors. The picture is from CHOOZ
 detector: ``region 1" contains 5-ton target material, ``region 2" protects the target from PMT radioactivity and
contains the $\protect\gamma$-ray from neutron capture, ``region 3" is used
to separate active cosmic-ray muon veto. This figure is taken from paper
\protect\cite{CHOOZ}.}
\label{figchoozDetector}
\end{figure}

\begin{figure}[tbp]
\begin{center}
\includegraphics*[0pt,0pt][580pt,454pt]{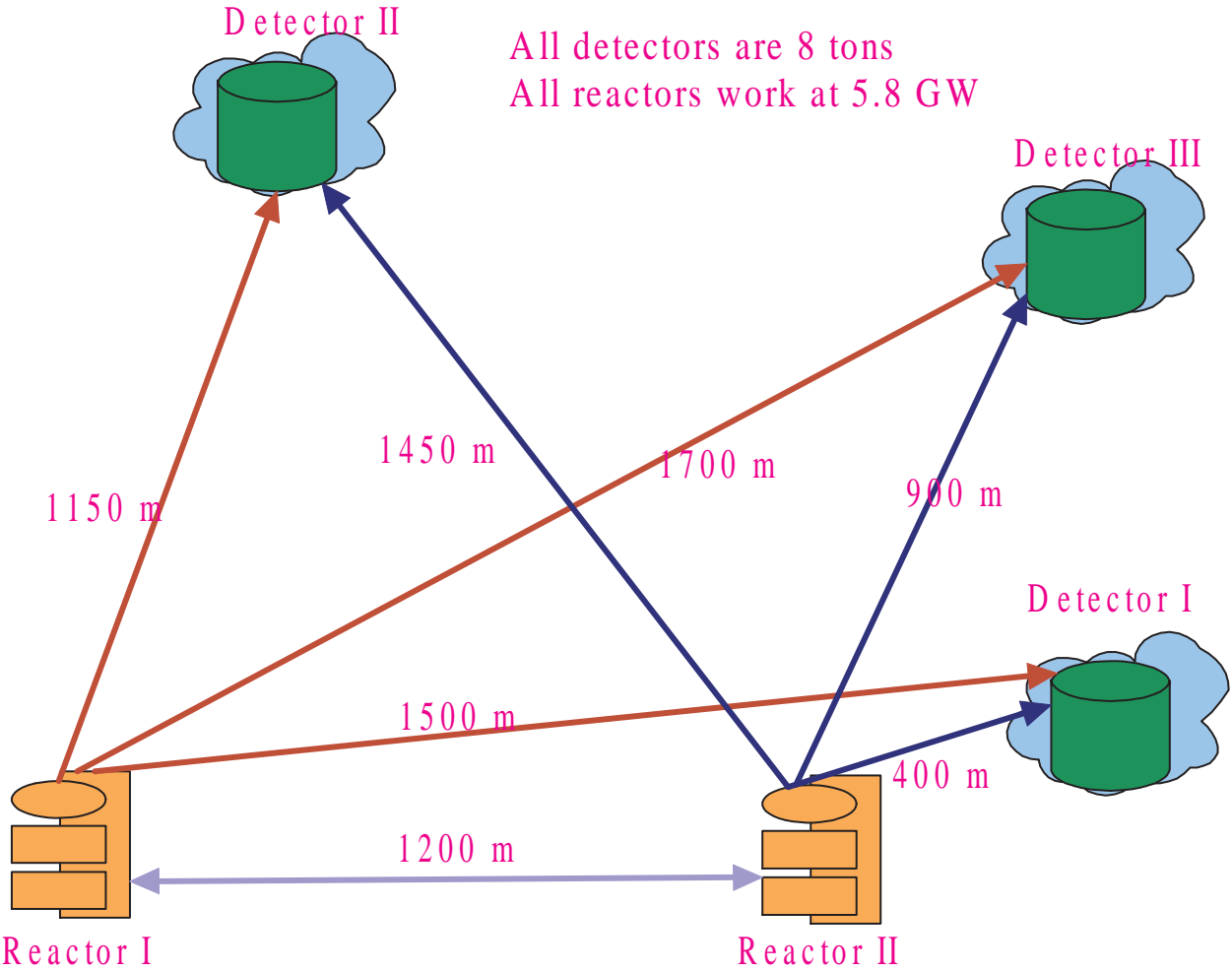} 
\end{center}
\caption{First scheme: we arrange three detectors in 
distances from the reactors in the range of $400m \sim 1700m$
unsymmetrically. }
\label{figArrange1}
\end{figure}

\begin{figure}[tbp]
\begin{center}
\includegraphics*[0pt,0pt][464pt,454pt]{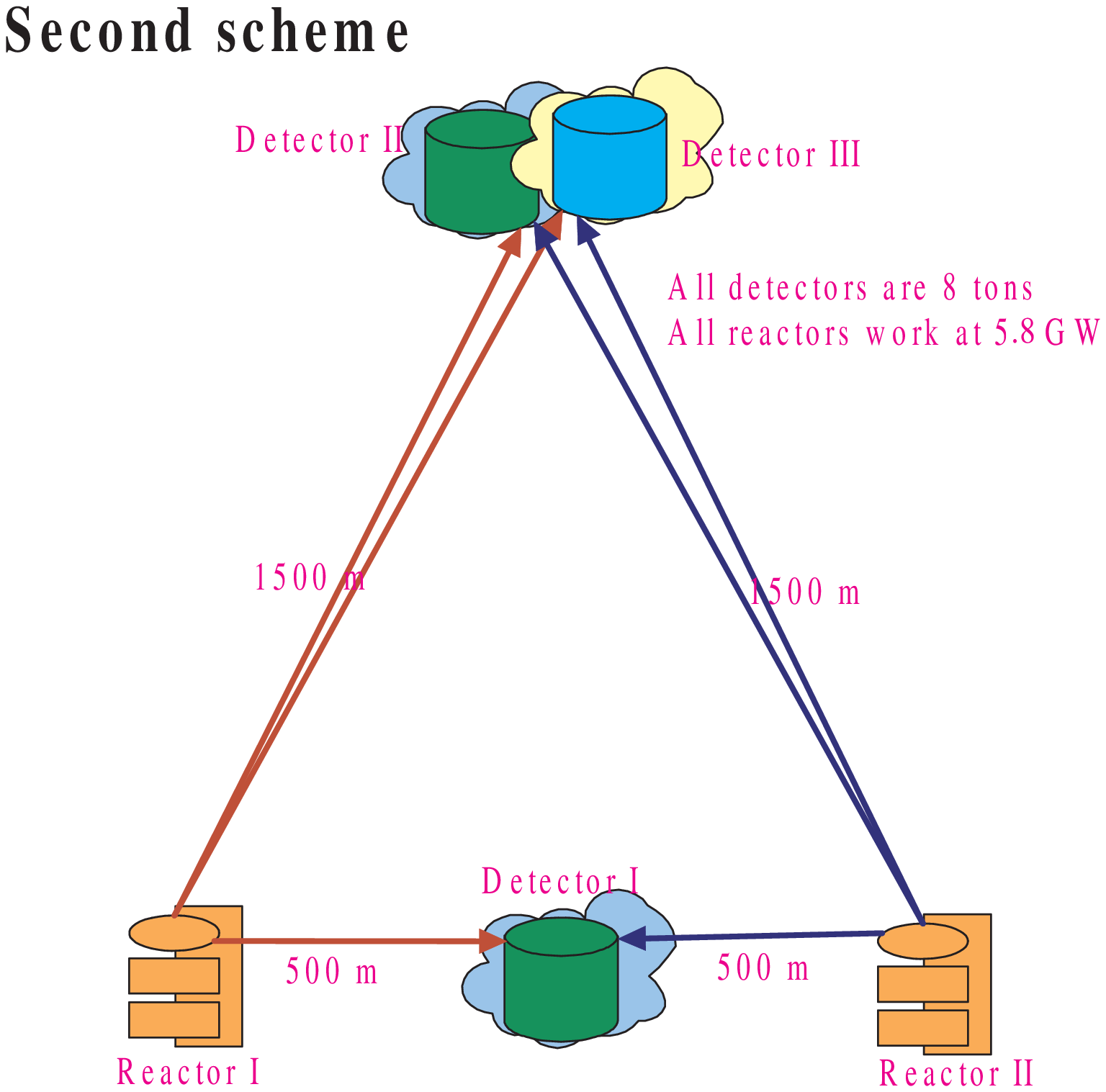} 
\end{center}
\caption{Second experiment scheme: three detectors
are placed symmetrically according to two reactors, one is on the
middle points, the other 
two are superposed on the perpendicular bisector equidistant 1500m from the
two reactors.}
\label{figArrange2}
\end{figure}

\begin{figure}[tbp]
\begin{center}
\includegraphics*[0pt,0pt][464pt,454pt]{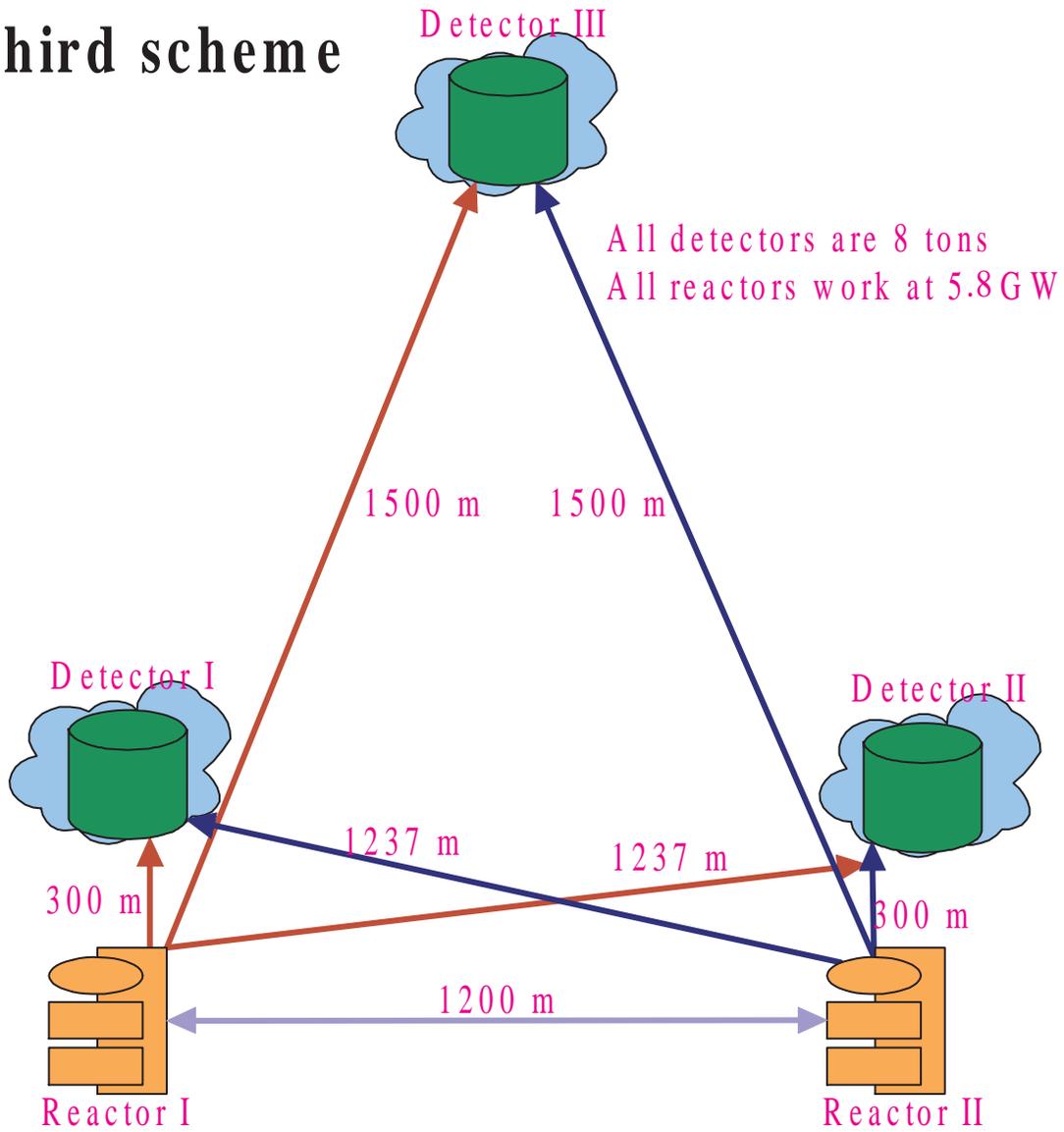} 
\end{center}
\caption{Third scheme: we arrange three detectors symmetrically
also, but from the near detectors, the two reactors are not equidistant.}
\label{figArrange3}
\end{figure}

\begin{figure}[tbp]
\begin{center}
\includegraphics*[0pt,0pt][490pt,500pt]{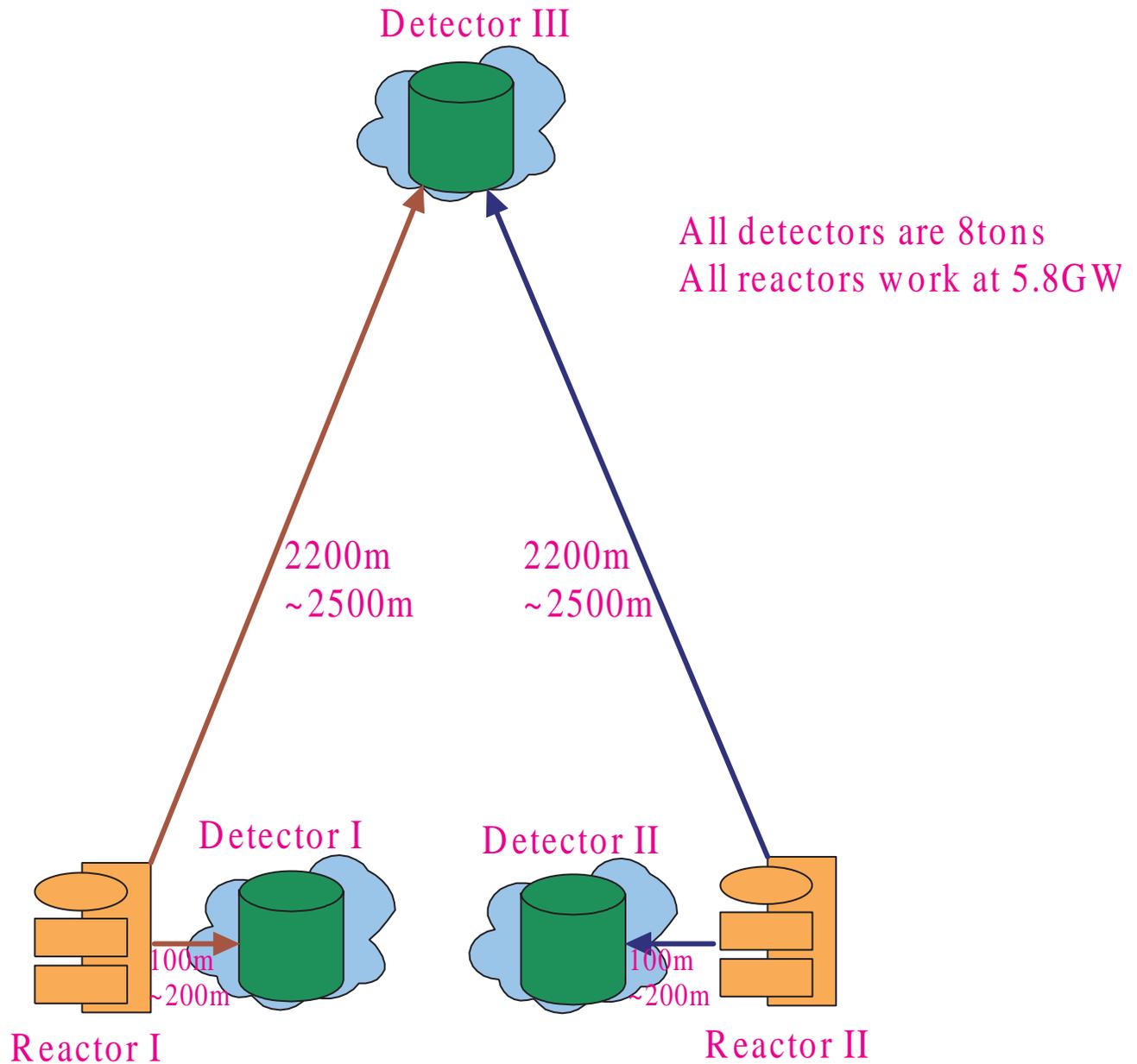}
\end{center}
\caption{Fourth scheme: similar to scheme 3, but two near detectors
  are put on the line between two reactors; and the distance of
  the far detector is enlarged to $2200\sim2500$ $m$.  }
\label{4scheme}
\end{figure}

\bigskip Three liquid scintillation calorimeter
detectors can be placed from distances between several hundreds meters
(near detectors) to about $2$ $km$ (far detectors) from the cores as shown in
figs. \ref{figArrange1} - \ref{4scheme}. 
High intensity and purity electron anti-neutrino flux from the reactor core is
detected via the inverse beta-decay reaction
eq. (\ref{inverse-decay}),  the signature is a delayed coincidence between the
prompt $e^{+}$ signal and the signal from the neutron capture, the target
material can be the Hydrogen-rich (free protons) paraffin-based liquid
scintillator loaded with Gadolinium, which is chosen due to its large
neutron capture cross section and to the high $\gamma $-ray energy released
after $n-$capture. The rock overburdens are able to reduce the external
cosmic ray muon flux by a fact of more than 300. To protect the detector
from natural radioactivity of the rock, the steel vessel should be
surrounded by $\sim 100cm$ of low radioactivity sand and covered by $\sim
15cm$ of cast iron. There are three concentric regions in a detector: as
shown in fig. \ref{figchoozDetector}: one is a central
8 tons working target in a transparent Plexiglas container filled with a $%
0.09\%$ Gd-loaded scintillator; the second should be an intermediate region,
equipped with hundreds eight-inch PMT's, used to protect the target from PMT
radioactivity and to contain the gamma rays from neutron capture; the third
can be an outer optically separated active cosmic-ray muon veto shield
equipped with two rings of tens eight-inch PMT's. All these arrangements
could significantly decrease the most dangerous background. Moreover, the
particular coincidence between the prompt positron signal and the delayed
neutron capture signal could significantly reduce the accidental background.
Rates from background can be suppressed down to $0.2\thicksim 0.3$ events per
day per ton, which corresponds to more than $60\%$ of a detection efficiency.

Besides the uncertainty in the efficiency, there are other uncertainties
contained in the experiment: the reaction cross section uncertainty from an
overall and conservative uncertainty on integral neutrino rate, $1.9\%$; the
number of target protons uncertainty ($0.8\%$), mainly because of the
difficulty in determining the hydrogen content; the overall precision on the
thermal power is claimed to be $0.7\%$; the uncertainty of the average
energy released per fission of the main fissile isotopes is $0.6\%$. The
five uncertainties presented above are all overall effective, and we could
combine them to a total effect with a value of $2.7\%$. However, as the
uncertainty of the energy scale ($1.1\%$) affects the experimental result
varying with respect to the energy bins, we have to specially consider
its influence.

\bigskip

It is supposed that only one of the two nucleon power plants would
run for the most of the total time of the experiment. This allow us to
know from which reactor the signal is generated. Otherwise,     
the detector should be desired to be able to distinguish the events
induced by the first reactor binary-core from the other. In the paper
\cite{nedirection}, a good determination of the anti-neutrino incoming
direction is discussed. It is based on
the neutron boost in the forward direction 
via the inverse beta-decay reaction, which is induced by the incident
neutrino; the kinetic energy of the neutron remains even after collisions
with protons inside the detector.

\section{ Three methods of Analysis }

\label{analyze_sec}

\bigskip 

Let us suppose that the experiment like DaYa-Bay will measure the positron spectrum in 
7 bins (from $E_{1}$ to $E_{7})$. For a mean reactor-detector distance 
$L_{k}$, the rates can be written as    
\begin{equation}
S_{k}(E,L_{k},\theta ,\Delta m^{2})=\frac{N_{p}}{4\pi L_{k}^{2}}\int
h(L,L_{k})\int \sigma (E_{\nu })S(E_{\nu })P(E_{\nu },L,\theta ,\Delta
m^{2})r(E_{e^{+}},E)\varepsilon (E_{e^{+}})dE_{\nu }dE_{e^{+}}dL
\end{equation}%
where

\begin{center}
\begin{tabular}{ll}
$E_{\nu},E_{e^+} $ & are neutrino energy and
positron energy respectively, \\ 
$N_p$ & is the total number of target protons \\ 
& in the Region I scintillator, \\ 
$\sigma(E_{\nu})$ & is the detection cross section, \\ 
$S(E_{\nu})$ & is the anti-neutrino spectrum, \\ 
$h(L,L_k)$ & is the spatial distribution function for \\ 
& the finite core and detector sizes, \\ 
$r(E_{e^+},E)$ & is the detector response function linking \\ 
& the visible energy $E$ and the real positron energy$E_{e^+}$ \\ 
$P(E_{\nu},L,\theta,\Delta~m^2)$ & is the two-flavor survival probability, \\ 
$\varepsilon(E_{e^+})$ & is the neutrino detection efficiency. \\ 
& 
\end{tabular}
\end{center}

The fissile isotope composition varies with respect to the working time,
because the 4 isotopes burn a bit differently. The expected number of
neutrino is a function of working time, reactor power and a constant
background. 
\begin{equation}
N_{i}^{j}=(B^{j}+W_{1j}\cdot Y_{1i}^{j}+W_{2i}\cdot Y_{2i}^{j})\cdot \Delta
t_{i}
\end{equation}%
where the index $i$ labels the run number or the working time information; $%
\Delta t_{i}$ is the corresponding live time interval; $j$ labels the
detector number; $B^{j}$ is the background rate, which is assumed to be a
constant with time in our study; ($W_{1i}$, $W_{2i}$) are the thermal powers
of the two binary-core reactors in GW and ($Y_{1i}$, $Y_{2i}$) are the
positron yields per GW per day induced by each reactor. Since we are
considering several detectors, it is convenient to factorize $Y_{ki}^{j}$
and $B^{j}$ into functions which separate the factors which are
independent of the detectors' size: 
\begin{equation}
Y_{ki}^{j}=(1+\eta _{ki}^{j})\cdot \frac{(1~km)^{2}}{(L_{k}^{j})^{2}}\cdot
X_{k}^{j}\cdot T^{j} ,
\end{equation}%
\begin{equation}
B_{i}^{j}=b_{i}^{j}\cdot T^{j} ,
\end{equation}%
where $k=1,2$ is the index of the reactors, associated with $T^{j}$ tons of
available detector material, $\eta _{ki}^{j}$ stands for corrections from
the reactor's differences on the fissile isotope composition and the positron
efficiency correction. The complications varying with time are all involved
in $\eta _{ki}^{j}$; this effect will be considered in real measurement and
 will not be taken into account in our simulations. $X_{k}^{j}$ represent
the positron events contribution in $1$ ton fresh detective material from a
fresh reactor located 1km away, with its thermal power equal to 1GW within
one day. In the following we compare the two positron spectra
$X_{k}^{j}$ which are the measured one and the oscillation expected
one. It can be parameterized by separating 
the oscillation term from the no-oscillation one: 
\begin{equation}
X_{k}^{j}(E_{l},L_{k}^{j},\theta ,\Delta m_{13}^{2})=X_{0}(E_{l})\overline{P}%
(E_{l},L_{k}^{j},\theta ,\Delta m_{13}^{2}) ,
\end{equation}%
where $X_{0}(E_{l})$ is the no-oscillation positron spectrum, unitary for
all the detectors in our definition, and $E_{l}$ label the visible energy
bins: 
\begin{equation}
X_{0}(E_{l})=\frac{n_{p}}{4\pi (1~km)^{2}}\int \sigma (E_{\nu })S(E_{\nu
})\int r(E_{e^{+}},E)\varepsilon (E_{e^{+}})dE_{e^{+}}dE_{\nu }  \label{Xint}
\end{equation}%
and $\overline{P}$ is the survival probability averaged over the energy bins
and the finite sizes of both the detector and the\ reactor core, 
\begin{equation}
\overline{P}(E_{l},L_{k}^{j},\theta ,\Delta m^{2})=\frac{\int
S_{k}(E,L_{k}^{j},\theta ,\Delta m^{2})dE}{\int S_{k}(E,L_{k}^{j},0,0)dE}
\end{equation}

With these definitions, we can begin to investigate our detector systems'
power. In the reactor neutrino oscillation experiment like DaYa-Bay,
we suggest that 
several detectors located at different places have no difference
between them, which means they have a same design; thus the systematic 
uncertainties are supposed to be the same in order to cancel out some
negative effects by comparison. In this paper, without losing generality, we
suppose that: three detectors have 8 tons of Gd-loaded scintillator 
each; two nuclear reactors each working at 5.8GW of its full thermal power.

To test an oscillation hypothesis ($\Delta m^{2}$, $\sin^{2}2\theta $) in our
experiment, we construct a $\chi ^{2}$ function including the 6 positron
spectra measured and the oscillation expected in a 42-element X array, as
follows: 
\begin{equation}
{\overrightarrow{X}=(\overrightarrow{X_{1}^{1}},\overrightarrow{X_{1}^{2}},%
\overrightarrow{X_{1}^{3}},\overrightarrow{X_{2}^{1}},\overrightarrow{%
X_{2}^{2}},\overrightarrow{X_{2}^{3}})} ,
\end{equation}%
\begin{equation}
\overrightarrow{X_{i}^{j}}%
=(X_{i}^{j}(E_{1}),X_{i}^{j}(E_{2}),X_{i}^{j}(E_{3}),X_{i}^{j}(E_{4}),X_{i}^{j}(E_{5}),X_{i}^{j}(E_{6}),X_{i}^{j}(E_{7})) ,
\end{equation}%
where the subscript $i$ is the reactor's number, and the superscript $j$ is
the detector's number. Combining the statistical variances with the
systematic uncertainties related to the neutrino spectra, the $42\times 42$
covariance matrix can be written in a compact form as follows: 
\begin{equation}
V_{ij}=\delta _{i,j}(\sigma _{i}^{2}+\widetilde{\sigma _{i}}^{2})+(\delta
_{i,j-21}+\delta _{i,j+21})\cdot \sigma _{12}^{(i)},~~(i,j=1,\cdot \cdot
\cdot 42)
\end{equation}%
where $\sigma _{i}$ are the statistical errors, $\widetilde{\sigma _{i}}$
are the corresponding systematic uncertainties, and $\sigma _{12}^{(i)}$ are
the statistical covariance of the reactor 1 and 2 yield contributions to the 
$i$-th measurement.

\noindent {\Large  A.} 
 
\noindent There are two kinds of uncertainties influencing all the
measurements which can't be ignored, the overall normalization
uncertainty which is $\sigma 
_{a}\thickapprox 2.7\%$, and the spectrum affective energy-scale calibration
uncertainty which is $\sigma _{g}=1.1\%$. We consider these two  
as in eq. (\ref{chiA}) to reach the minimum $\chi _{A}^{2}$ for
a set of 
proposed oscillation parameters: 
\begin{eqnarray}
\chi _{A}^{2}(\theta ,\Delta m^{2}) &=&\min_{a,g}\left[ \sum_{i,j}(\hat{X}^{i}-a%
\cdot \overline{\hat{X}}^{i}(gE,\theta ,\Delta m^{2}))\left( V^{-1}\right)
_{i,j}(\hat{X}^{j}-a\cdot \overline{\hat{X}}^{j}(gE,\theta ,\Delta m^{2}))\right.  
\notag \\
&&\left. +\left( \frac{a-1}{\sigma _{a}}\right) ^{2}+\left( \frac{g-1}{%
\sigma _{g}}\right) ^{2}\right],   \label{chiA}
\end{eqnarray}
where $\hat{X}^{i}$ stands for measured or simulated one, and
$\overline{\hat{X}}^{i}(gE,\theta ,\Delta m^{2})$ for oscillation
expected one. Here we have taken into account the statistic
uncertainty, so we add up a hat on  $X^{i}$. The statistic error
is the square root of the event number which depends on the detector
size, the reactor power and the experiment's life time.        
This method uses all the experimental information available, and directly
depends on the correct determination of the systematic uncertainties. Such
uncertainties could be reduced by measuring the positron spectrum with a
detector near the reactors. When a product of the reactor power, the detector
size and the working time is large enough to reduce the statistical
uncertainty, we could constrain this overall normalization
coefficient and the energy-scale calibration precisely.

\noindent {\Large  B.} 

\noindent  We compare the ratio of the two reactors'
contribution in another way.
Since the expected spectra are the same for both reactors in the case of
no-oscillation, the ratio reduces to the ratio of the average survival
probabilities. Detectors are assumed to be the same, the systematic
uncertainties will be canceled out or almost canceled out and the remaining
uncertainty we should take into account is just the statistical uncertainty.
So we can construct $\chi^2_B$ by 
\begin{eqnarray}
R^j_k=\frac{\hat{X}^j_1(E_k)}{\hat{X}^j_2(E_k)},~~~~~~\delta R^j_k=R^j_k\cdot \left(%
\frac{\delta \hat{X}^j_1(E_k)}{\hat{X}^j_1(E_k)}+\frac{\delta \hat{X}^j_2(E_k)}{\hat{X}^j_2(E_k)}%
\right),  \notag
\end{eqnarray}
\begin{eqnarray}
\chi^2_B(\theta,\Delta m^2)=\sum_{j,k}\left(\frac{R^j_k-R^j_k(\theta,\Delta
m^2)}{\delta R^j_k }\right)^2
\end{eqnarray}

$\chi _{B}^{2}$ is a good quantity to judge an oscillation result from
no-oscillation one in the 
case where the distances to every reactor for a detector are quite
different. In the CHOOZ experiment the distance difference is just $116.7$
$m$, so
that the $\chi _{B}^{2}$ is less powerful than $\chi _{A}^{2}$ \cite{CHOOZ}.
In our experiment design, the near detectors are expected to be placed
at positions 
which give a larger distance difference; thus we can expect $\chi
_{B}^{2}$ is more powerful.

\noindent {\Large  C.} 

\noindent  There is an intermediate analysis approach between the
$\chi _{A}^{2}$ and  $\chi _{B}^{2}$. It uses
the shape of the positron spectrum, while leaving the absolute normalization
free. Similar to approach A, this approach fits the two uncertainties, but
leaving the normalization parameter unconstrained, that is: 
\begin{equation*}
\chi _{C}^{2}(\theta ,\Delta m^{2})=\min_{a,g}\left[ \sum (\hat{X}^{i}-a\cdot 
\overline{\hat{X}}^{i}(gE,\theta ,\Delta m^{2}))\left( V^{-1}\right)
_{i,j}(\hat{X}^{j}-a\cdot \overline{\hat{X}}^{j}(gE,\theta ,\Delta m^{2}))+\left( \frac{%
g-1}{\sigma _{g}}\right) ^{2}\right] 
\end{equation*}

In the following, we will use these three analysis methods to constrain the
oscillation parameters for the possible reactors and detectors systems. With
these results, we are able to judge which kind of system and the
corresponding approach is the most powerful one.

\section{ Examining Arrangements and Simulation Results }

Comparing with the CHOOZ experiment, DaYa-Bay has more powerful
reactors, longer working time, more larger detectors.  Let us suppose
the nuclear power plants work at its full thermal power 5.8(GW), the detector's
available mass is 8 tons, and the experiment life time is set to be one
year in our first simulation. We assume we had known the signal's
corresponding reactor individually.    

\begin{figure}[tbp]
\begin{center}
\includegraphics*[bb=30 5 500 700,height=1.\textwidth]{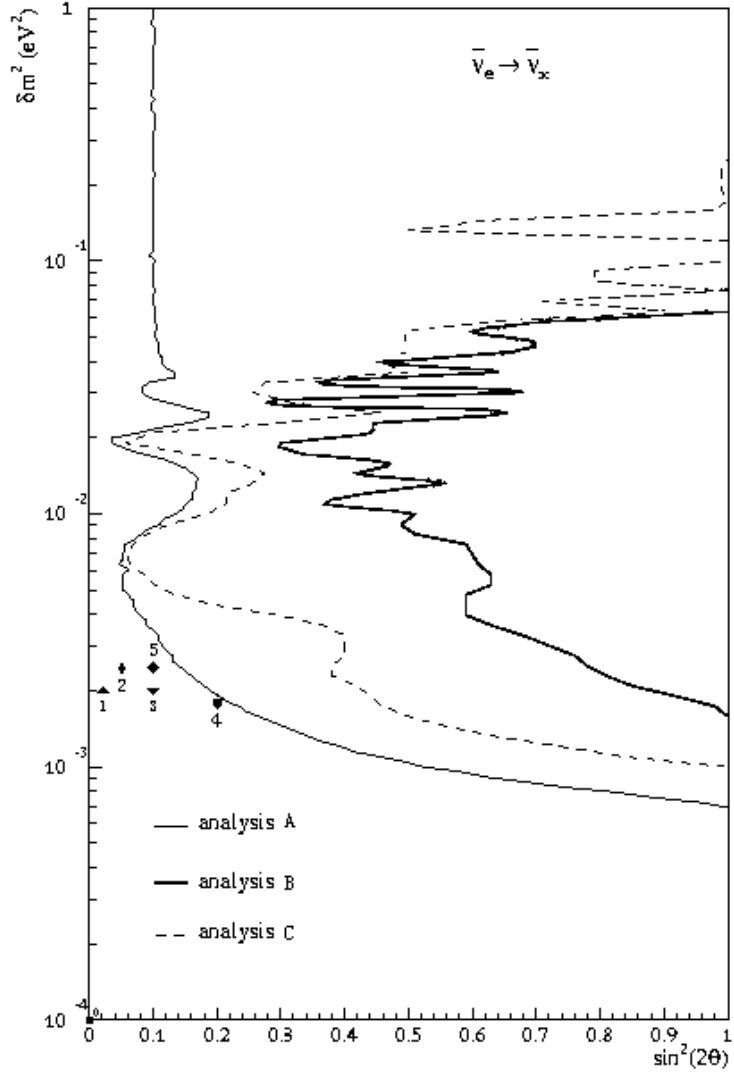}
\end{center}
\caption{ Six points whose coordinates are considered as real physics
  parameters for the first 3 schemes' simulations, in CHOOZ's excluded
  region.   
First ponit represents no-oscillation, labelled as ``0''. The other
  five points below
will cause oscillation effect; their parameters ($\sin ^22\protect\theta%
_{13}$, $\Delta m^2$ [eV$^2$]  ) are 1$\rightarrow$(0.02, $2.0\times10^{-3}$), 2$%
\rightarrow$(0.05, $2.5\times10^{-3}$), 3$\rightarrow$(0.1, $%
2.0\times10^{-3} $), 4$\rightarrow$(0.2, $1.7\times10^{-3}$), 5$\rightarrow$%
(0.1, $2.5\times10^{-3}$).  }
\label{figchoozcont}
\end{figure}

1) In our first scheme, three detectors with distances from the
reactors in the range of $400m\sim 1700m$ are arranged asymmetrically
as plotted in fig.  \ref{figArrange1}. Every detector can give two
different distances with its 
corresponding oscillation spectra. This scheme is
based on an idea: the allowed region of $\Delta m^{2}$ ($=\Delta
m_{atm}^{2})$ given by the atmospheric data is still too big to determine the
best position for all detectors; in order to consider all the possible 
$\Delta m^{2}$, we try to place the detectors at different positions
which can provide more numbers of different distances. We test our system with
some possible parameters chosen from the CHOOZ's indistinguishable
region; the points in that parameter region are shown in 
fig. \ref{figchoozcont}. We simulate the experiment by using the 
values of those points one by one as the values of true physics
parameters. Like in previous discussions, we assume each possible 
oscillation hypothesis to construct the excluded/allowed region at
different confidence levels \cite{FCbelt}.

\begin{figure}[tbp]
\begin{center}
\includegraphics*[49pt,190pt][570pt,620pt]{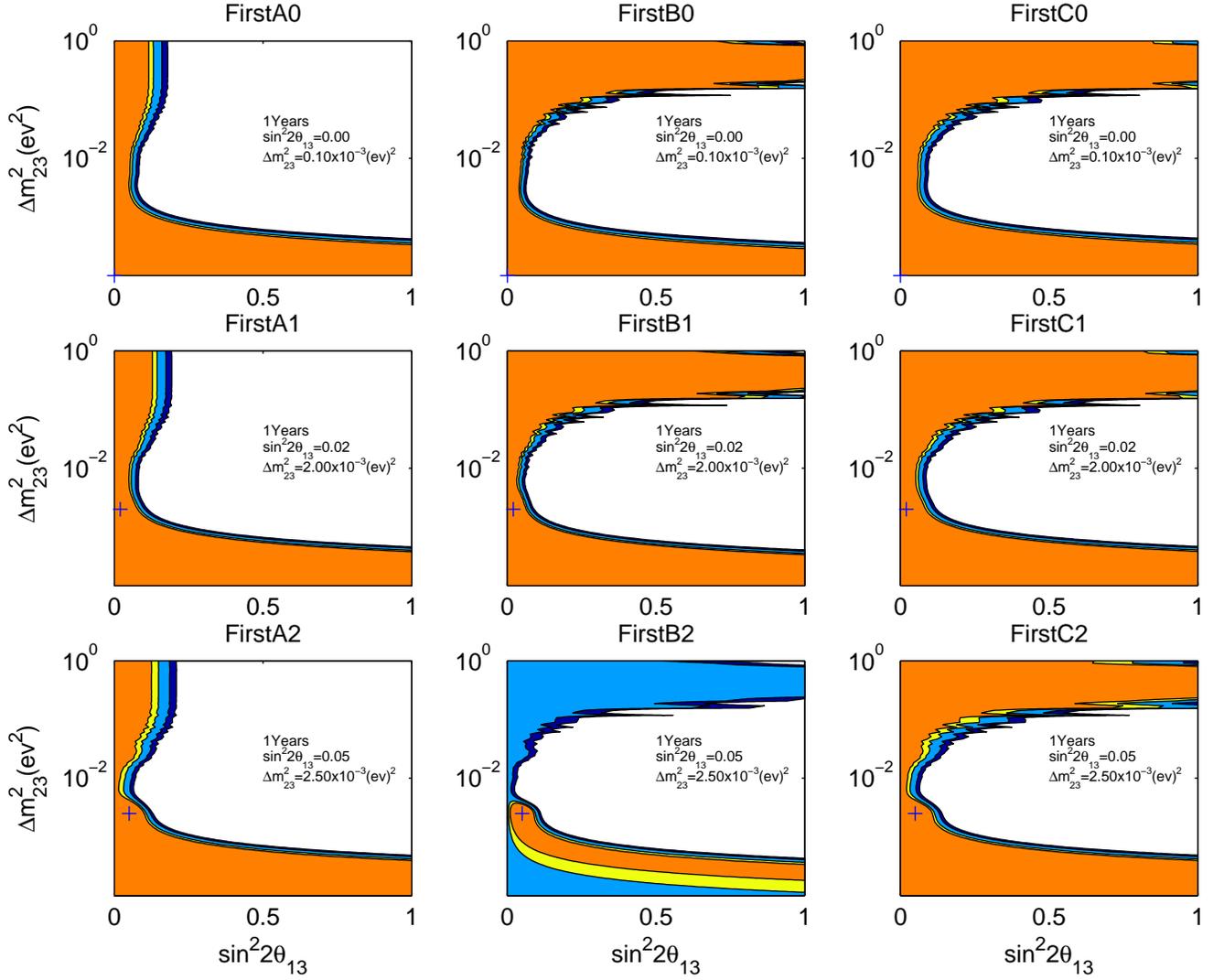}
\end{center}
\caption{ The allowed regions at different confidence levels for the
  first scheme. The 
labels ``First", ``Second", ``Third" stand for different schemes; ``A",
``B", ``C" 
for different analysis methods, and ``0 $\sim$ 2" for the simulation input
parameters with their value presented in the figure, in a plot,
it is denoted by a plus sign. From this point to the exclusion area,
the confidence levels of 
the four regions are $90\%, 95\%, 99\%, 99.73\%$. }
\label{Myfig1}
\end{figure}

\begin{figure}[tbp]
\begin{center}
\includegraphics*[49pt,190pt][570pt,620pt]{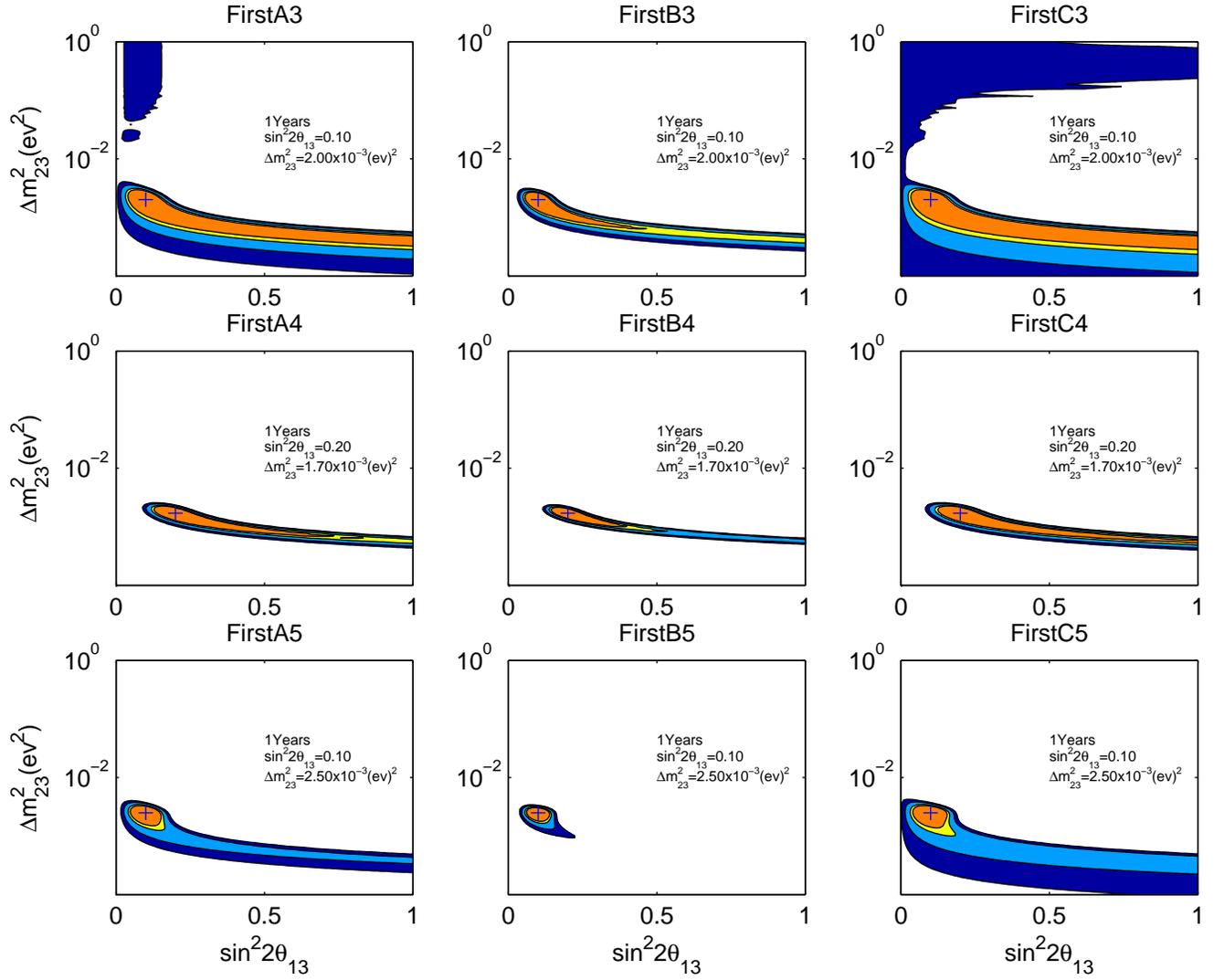}
\end{center}
\caption{Continue plots of fig. \ref{Myfig1}}
\label{Myfig2}
\end{figure}

The  Monte Carlo simulation results are presented in
 figs. \ref{Myfig1}, \ref{Myfig2}, 
 using three analysis approaches for the six representational possible "true
value" points  as signed in fig. \ref{figchoozcont}. For our first
detectors' arrangement, we can see that approach B is the most
 powerful one while 
A is less sensitive and C is the lowest. 
In the following discussion, we will not present our
approach C results since A is similar and better. Using analysis
approach B in First scheme, the experiment like DaYa-Bay is able to find an
oscillation result for $\sin ^{2}2\theta _{13}$ bigger than 0.05 (at
more than $90\%$ C.L. ), if
$\Delta m_{23}^{2}$ is at the region of $\sim ~2.5\times 
10^{-3}\mathtt{eV}^{2}$; and the oscillation parameters can be
even constrained to a small allowed region (fig. \ref{Myfig2}) for bigger $\sin
^{2}2\theta _{13}$ (as the last three points we simulated
with). It is interesting that one can 
bound $\sin ^{2}2\theta _{13}$ to be bigger than 0.05 at $95\%$ C.L.,
if the true parameter is located near the point ($\sin ^{2}2\theta
_{13}=0.1$, $ 
\Delta m^{2}=2.0\times 10^{-3}$eV$^{2}$); and the best restriction on $\sin
^{2}2\theta _{13}$ is reached when the simulated parameters is located near ($
\sin ^{2}2\theta _{13}=0.1$, $\Delta m^{2}=2.5\times 10^{-3}$eV$^{2}$);
there we could use this system combined with analysis B to constrain the
oscillation mixing angle and the mass square difference to a precise region
as we present in the subplot (3rd row, 2nd column) named as 'FirstB5' of fig.
 \ref{Myfig2}.

2) In the second scheme\cite{yfWangppt}, fig. \ref{figArrange2}, the three
detectors are arranged symmetrically from the reactors, one is in the
middle of two reactors, the other two are superposed on the
perpendicular bisector 
equidistant 1500m from the two reactors. Since the two reactors are
symmetric to every detector, approach B is disabled. Approach A can also
distinguish $\sin^{2}2\theta _{13}$ bigger than 0.05 if $\Delta m_{23}^{2}
$ is near $2.5\times 10^{-3}eV^{2}$; but it can only constrain $%
sin^{2}2\theta _{13}$ at $95\%$ C.L. and the maximum mixing is not excluded
at $99\%$ C.L. even at the most sensitive oscillation points, if by
the DaYa-Bay experiment itself.

\begin{figure}[tbp]
\begin{center}
\includegraphics*[49pt,190pt][570pt,620pt]{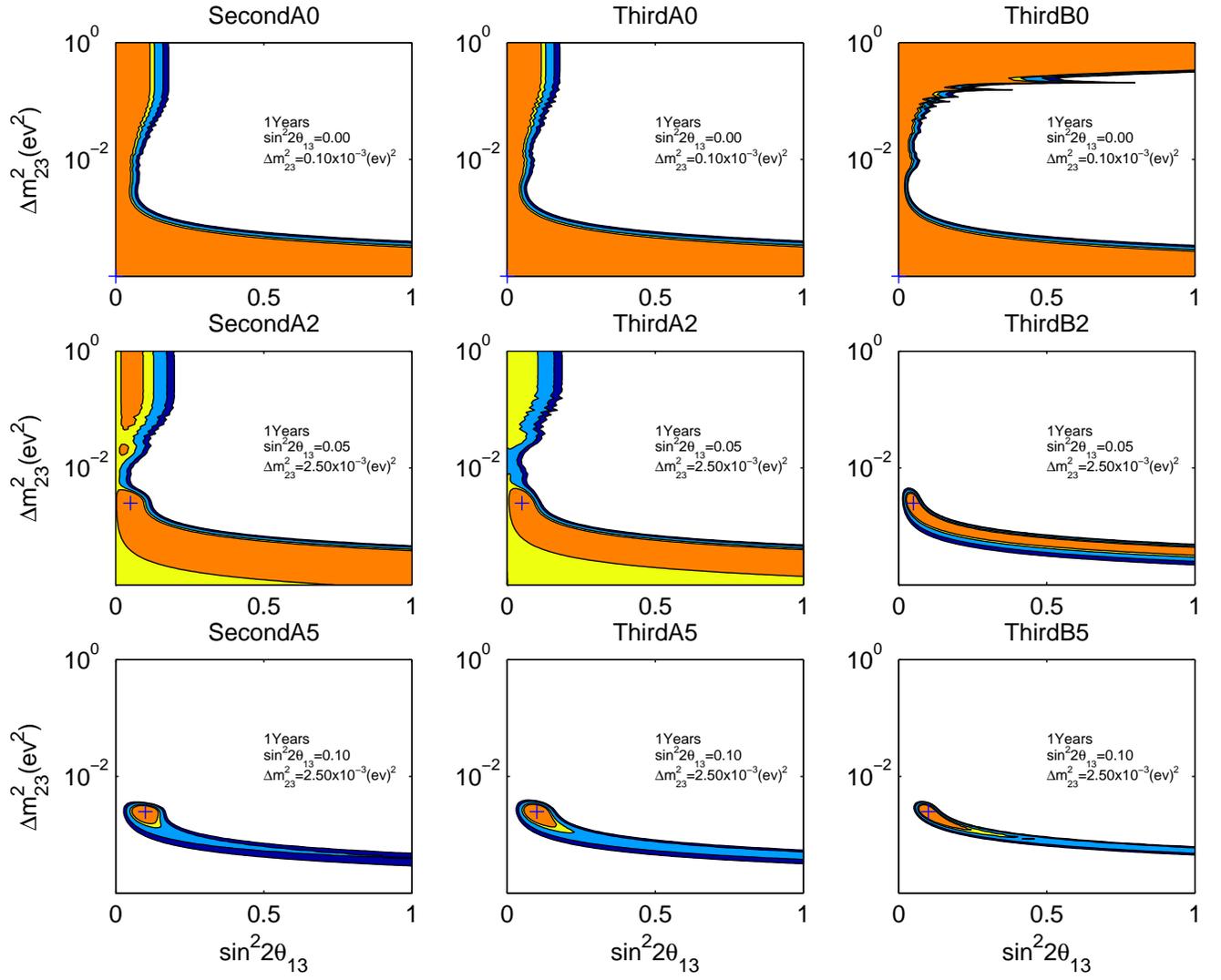}
\end{center}
\caption{ The allowed regions (shadows) of our simulation for second and third
schemes, same notations as previous plots.}
\label{Myfig3}
\end{figure}

3) In the third scheme \cite{yfWangppt}, fig. \ref{figArrange3}, detectors are
also disposed symmetrically, but from the near detectors, the two reactors
are not equidistant, so the analysis approach B is partly
available. For one year's operation of the experiment, this scheme
shows a very clear oscillation signal for 
$\sin^2 2\theta_{13}$ larger than $0.05$ at 99.73$\%$ C.L. (2nd
row, 3rd column in fig. \ref{Myfig3}).

\begin{figure}[tbp]
\begin{center}
\includegraphics*[49pt,190pt][570pt,620pt]{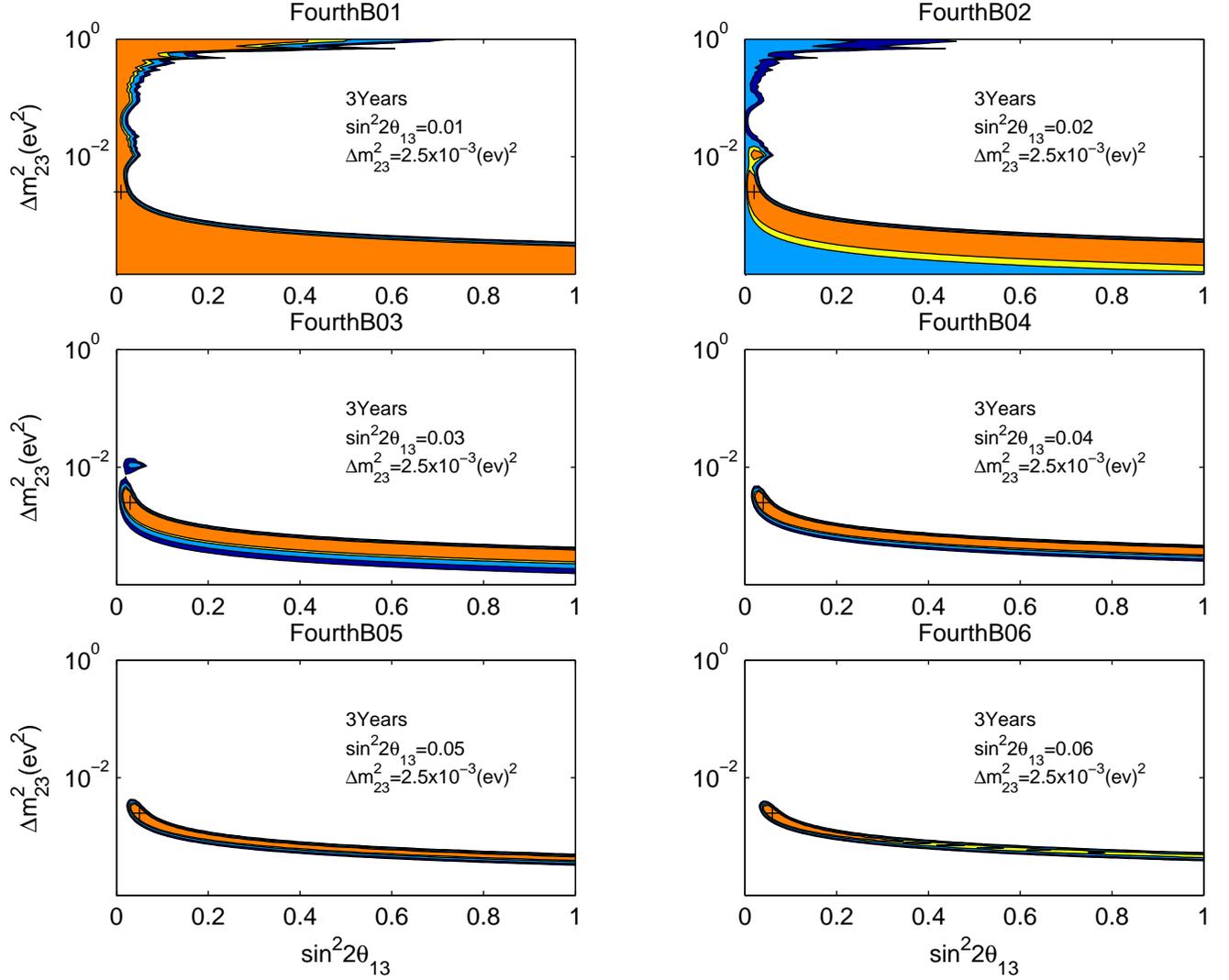}
\end{center}
\caption{Similar to previous figures, these are results for the
Fourth scheme with 3 years' experiment. The title ''FourthB0x'' stands
for $\protect\chi^2_B $ analysis of the simulation with the parameters
in the figure.}
\label{3years}
\end{figure}

4) Scheme three is quite good since it gives a higher confidence
level than First and Second schemes if there is a positive result for a
non zero $\sin^2 2\theta_{13}$. However it is still possible to improve it.
As a conclusion, in this paper, for detectors' locations, we suggest a
 possible setting for the 
experiment like DaYa-Bay, which is called the Fourth scheme: an extension of the Third
scheme, which is to put a 8-ton 
detector $2200m \sim 2500m $ symmetrically away from two reactors; and
put the other two 8-ton detectors more close to their corresponding
reactors respectively, almost at a $100m \sim 200m$ distance; they
are located on the line between the two reactors (see
fig. \ref{4scheme}). The reason for a  $2200m \sim 2500m$-detector is
based on the most sensitive oscillation zone with respect to the range
of present $\Delta m^2_{31}$, taking into account the whole energy
spectrum effect of reactor neutrinos. The best way for the two
near detectors to easily distinguish which reactor a neutrino signal
comes from, is to put the other two detectors on the inner line
between two reactors; thus two different neutrino sources are from two
opposite directions. We use the $\chi^2_B$ method to analyze Monte
Carlo results, in which the far detector's data is not taken into
account since this detector is symmetric to two reactors; this doesn't
affect our major statement about the discovery potential of this
experiment. However, the far detector is important in a real data
analysis: using the single $\chi^2_A$ method it will exclude big-$\sin^2
2\theta_{13}$ region, and 
give a precise allowed region in a combined analysis of $\chi^2_A$
and $\chi^2_B$; it can also be used to implement much smaller
systematic errors in other different analysis method not discussed in
this paper. During 3 years of running, the experiment like
DaYa-Bay is able to
discover a non zero $\theta_{13}$  if $\sin^2 2\theta_{13} \ge 0.02$ at 95$\%$
C.L. (fig. \ref{3years}), and at 99.73$\%$ C.L. for $\sin^2 2\theta_{13}
\ge 0.03$. This result is able to exclude the most general SO(10)
GUTs, an inspiring Grand Unification candidate, if nature doesn't
choose it.

 As we have seen, the experiment like DaYa-Bay described in this paper is
 simple, has no technical difficulty, and could have been realized even several
years ago; but with the optimization to the detectors position, we could get
a more precise result ($\sin^{2}2\theta _{13} \sim 0.02$) than what we
 ever had \cite{CHOOZ}. Moreover, it is 
possible that some improvements in the technique, more detectors and advanced
methods could be used \cite{yfWangppt} to have more precise results. It is
promising that, to a greater extent, this experiment could reach a
precision of $\sin^{2}2\theta _{13} \sim 0.01$. If the detectors are
 constructed as movable 
objects, they can be used to measure the solar neutrino parameters
($\sin^2 2\theta_{12}, \Delta m^2_{12}$) in the next phase. 

\section{Conclusion}

The reactor neutrino experiment like DaYa-Bay offers an opportunity to
discover a non zero $\theta_{13}$, another crucial step
in particle physics after solving the solar
neutrino problem. We arrange four schemes for the three 8-ton detectors' 
locations, and select the fourth scheme as our suggestion for the
experiment. In the First scheme, with respect to two reactors, we
place 
three detectors as asymmetric as possible in the distance range of
$400m$ to $1700m$, in order to have more oscillation distances. We
relax a systematic uncertainty to totally a few percent (much bigger
than one percent that is a possible but difficult achievement), which
is already reached by CHOOZ's technology. During three years of data
taking, the simulation result shows that a discovery ability of this scheme
is $\sin^2 2\theta_{13} \ge 0.03$; while one year operation can
give for $\sin^2 2\theta_{13}$ a limit of $0.05$. The Second scheme is to
place two 8-ton detectors at $1500m$ in the same place, while a third $500m$
symmetrically away from the two reactors. The Third scheme is to put a 8-ton
detector $1500m$ symmetrically away from the reactors; for the other
two, both of them 
are $300m$ away from one reactor and $1237m$ from the other reactor,
as shown in fig. \ref{figArrange3}. Both of these schemes are able
to reach 
a limit of $\sin^2 2\theta_{13}$ at $0.05$, during one year of
data taking; moreover, the Third scheme gets this sensitivity with
the highest confidence level of $3\sigma$ (fig. \ref{Myfig3}, 2nd
row, 3rd column). We conclude that for a discovery potential, the Third
scheme is a bit better; for a precise measurement after discovering a
non-vanishing $\theta_{13}$, the First scheme is better. Furthermore,
we suggest as the best possible location of detectors for the
experiment like DaYa-Bay, the Fourth scheme:
an extension of the Third scheme, which is to put a 8-ton
detector $2200m \sim 2500m $ symmetrically away from two reactors; and
put the other two 8-ton detectors more close to their corresponding
reactors respectively, at about $100m \sim 200m$ distance; and they
are located on the line between the two reactors. The Fourth
scheme will be able to discover a $\sin^2 2\theta_{13} \ge 0.02$ at 
2$\sigma$ level (fig. \ref{3years}, 1st row, 2nd column), for 3 years
of running the experiment like DaYa-Bay. 
  With improvement in technology and better budget, the sensitivity to
$\sin^2 2\theta_{13}$ can be even better. 

\noindent{\large \textbf{Acknowledgments:}} 
This work has been done quite a long time ago. However the text and conclusion
don't need to change even after the appearance of the double CHOOZ's paper, 

One of the authors, Q.Y.L., would like to thank 
A. Yu. Smirnov for reading of the paper and useful suggestions, and
the Abdus Salam 
International Centre for Theoretical Physics for hospitality. This
work is supported in part by the National Nature 
Science Foundation of China.

\newpage

\section{Appendix: The minimization method in approaches $\protect\chi^2_A$
and $\protect\chi^2_C$}

In section \ref{analyze_sec}, we defined in Eq. (\ref{chiA}) that the $
\chi^2_A$ is the minimum of the fitting of $a,g$ for the oscillation
parameters. Because the energy-scale calibration factor $g$ is involved in
the integral for $X$, such as in Eq. (\ref{Xint}), it is troublesome to get an
analytical expression for the minimization, and the numerical computation is
also insufferable if scanning the parameters plane is necessary. In
order to get a precise minimum quickly, we assume that: 
\begin{eqnarray}
\frac{X^i(g,E_l,\theta,\Delta m^2)-X^i(g=1,E_l,\theta,\Delta m^2)}{%
X^i(g=1,E_l,\theta,\Delta m^2)}=f(E_l)\cdot (g-1)
\end{eqnarray}
where $f(E_l)$ is the ratio of the two differences, gained from the
no-oscillation case, since the oscillation effect is very small even though it
can be detected in our system, these ratios are suitable for the oscillation
case. We check the linear assumption at every energy bin, it holds when $g$
changes near $1.0$ in the range of $5\%$. This property is good
enough when the uncertainty of $g$ is just $1.1\%$. For convenience, we
rewrite Eq. (\ref{chiA}) as 
\begin{eqnarray}
\chi^2_A=\min_{a,g}W(a,g)
\end{eqnarray}
where 
\begin{eqnarray}
W(a,g)=\sum_{i,j}\left((X^i-a
\overline{X}^i(gE))\left(V^{-1}\right)_{i,j}(X^j-a \overline{X}^j(gE))%
\right)+\left(\frac{a-1}{\sigma_a }\right)^2+\left(\frac{g-1}{\sigma_g}%
\right)^2
\end{eqnarray}
Here we have omitted the oscillation parameters in the bracket, keeping in
mind that $X$ followed by bracket is the oscillation prediction, while
single $X$ is the experimental or Monte Carlo simulated result. Using the
linear assumption, we can get $W(a,g)$'s partial derivatives $%
\partial_aW(a,g),\partial_gW(a,g)$ easily, moreover, the numerical
computation is just the multiplication of vectors and matrices. 
\begin{eqnarray}
\frac{\partial W(a,g)}{\partial a}= 2\sum_{i,j}\left[-X^i(g=1)\left(V^{-1}%
\right)_{i,j}X^j-(g-1)(fX)^i\left(V^{-1}\right)_{i,j}X^j \right.  \notag \\
\left.
+aX^i(g=1)\left(V^{-1}\right)_{i,j}X^j(g=1)+2a(g-1)(fX)^i\left(V^{-1}%
\right)_{i,j}X^j(g=1) \right.  \notag \\
\left. +a(g-1)^2(fX)^i\left(V^{-1}\right)_{i,j}(fX)^j \right]-2\frac{a-1}{%
(\sigma_a)^2}
\end{eqnarray}

\begin{eqnarray}
\frac{\partial W(a,g)}{\partial g}=-2\sum_{i,j}\left[a(fX)^i\left(V^{-1}%
\right)_{i,j}X^j-a^2(fX)^i\left(V^{-1}\right)_{i,j}X(g=1)^j \right.  \notag
\\
\left. -a^2(g-1)(fX)^i\left(V^{-1}\right)_{i,j}(fX)^j \right]-2\frac{g-1}{%
(\sigma_g)^2}
\end{eqnarray}
where $(fX)^i$ is $f(E_l)\cdot X^i(g=1,E_l)$. Starting from $(a=1,~g=1)$,
driven by $(-\partial_aW(a,g),-\partial_gW(a,g))$ at this point, and using
appropriate iterative step length, the minimization of $W(a,g)$ for an
oscillation hypothesis is quickly achieved.

\newpage 

\end{document}